\documentclass[aps,prb,showpacs,twocolumn,epsfig]{revtex4}
\usepackage{amssymb}
\usepackage{amsbsy}
\usepackage{amsmath}
\usepackage{epsfig}

\begin{document}

\title{Unconventional scanning tunneling conductance spectra for graphene}

\author{K. Saha $^{(1)}$, I. Paul $^{(2),(3)}$ and K. Sengupta $^{(1)}$}

\affiliation{$^{(1)}$ Theoretical Physics Division, Indian
Association for the Cultivation of Sciences, Kolkata-700032, India.\\
$^{(2)}$ Institut N\'{e}el, CNRS/UJF, 25 Avenue des Martyrs, BP 166,
38042 Grenoble, France.\\
$^{(3)}$ Institut Laue-Langevin, 6 rue Jules Horowitz, BP 156,
38042 Grenoble, France
}

\date{\today}

\begin{abstract}

We compute the tunneling conductance of graphene as measured by a
scanning tunneling microscope (STM) with a normal/superconducting
tip. We demonstrate that for undoped graphene with zero Fermi
energy, the {\it first derivative} of the tunneling conductance with
respect to the applied voltage is proportional to the density of
states of the STM tip. We also show that the shape of the STM
spectra for graphene doped with impurities depends qualitatively on
the position of the impurity atom in the graphene matrix and relate
this unconventional phenomenon to the pseudopsin symmetry of the
Dirac quasiparticles in graphene. We suggest experiments to test our
theory.

\end{abstract}


\maketitle

\section{Introduction}

The low energy quasiparticles of graphene around $K$ and $K'$ Fermi
points have Dirac-type properties \cite{neto1}. In particular, the
pseudospin of these quasiparticles around $K$($K'$) points along
(opposite to) their direction of motion. The presence of such
Dirac-type quasiparticles with definite helicity leads to a number
of unusual electronic properties in graphene
\cite{shar1,nov2,beenakker1,sengupta1}. Recently, the influence of
such Dirac quasiparticles on properties of graphene doped with
magnetic/non-magnetic impurities have attracted theoretical and
experimental
attention\cite{baskaran1,neto2,sensor1,hari1,peres1,stmgra1}.
However, the recent experimental observation of dependence of STM
tunneling spectra on the position of the impurity in the graphene
matrix in Ref.\ \onlinecite{hari1} lacks a theoretical explanation
even at a qualitative level \cite{peres1}.

Scanning tunneling microscopes (STM) are extremely useful probes for
studying properties of two or quasi-two dimensional materials
\cite{stmgra1,davis1}. Studying electronic properties of a sample
with STM typically involves measurement of the tunneling conductance
$G(V)$ for a given applied voltage $V$. The tunneling conductances
measured in these experiments have also been studied theoretically
for conventional metallic systems and are known to exhibit Fano
resonances at zero bias voltage in the presence of impurities
\cite{madhavan1,wingreen1}. The application and utility of this
experimental technique, with superconducting STM tips, have also been
discussed in the literature for conventional systems \cite{stmsc}.
However, tunneling spectroscopy of graphene using superconducting
STM tips remains to be studied both experimentally and
theoretically.

In this work, we compute the STM response of doped graphene and
demonstrate that the STM spectra has several unconventional
features. For undoped graphene with Fermi energy $E_F=0$, the
derivative of the STM tunneling conductance ($G$) with respect to
the applied voltage ($dG/dV$) reflects the density of states (DOS)
of the STM tip ($\rho_t$), ${\it i.e.}$, $dG/dV \sim +(-) \rho_t$
for $V>(<)0$. By tuning $E_F$, one can interpolate between this
unconventional $\rho_t \sim \pm dG/dV$ and the conventional $\rho_t
\sim G$ (seen for $E_F \gg eV$) behaviors. Further, we find that for
superconducting STM tips with energy gap $\Delta_0$, $G\, (dG/dV)$
displays a cusp (discontinuity) at $eV=-E_F-\Delta_0$ as a signature
of the Dirac point which should be experimentally observable in
graphene with small $E_F$ where the regime $eV > E_F$ can be easily
accessed. For impurity doped graphene with large $E_F$, experiments
in Ref.\ \onlinecite{hari1} have seen that the tunneling
conductance, as measured by a metallic STM tip, depends
qualitatively on the position of the impurity in the graphene
matrix. For impurity atoms atop the hexagon center, the zero-bias
tunneling conductance shows a peak; for those atop a graphene site,
it shows a dip. We provide a qualitative theoretical explanation of
this phenomenon and show that this unconventional behavior is a
consequence of conservation/breaking of pseudospin symmetry of the
Dirac quasiparticles by the impurity. We also predict that tuning
$E_F$ to zero by a gate voltage would not lead to qualitative change
in shape of the conductance spectra when the impurity is atop the
hexagon center; for impurity atop a site, the tunneling conductance
would change from a dip to a peak via an antiresonance.

The organization of the rest of the paper is as follows. In Sec.\
\ref{secfor}, we present the derivation of the tunneling current.
This is followed by Sec.\ \ref{secresult} where we present our main
results. Finally we conclude in Sec.\ \ref{seccon}.

\begin{figure}
\rotatebox{0}{\includegraphics*[width=0.95 \linewidth]{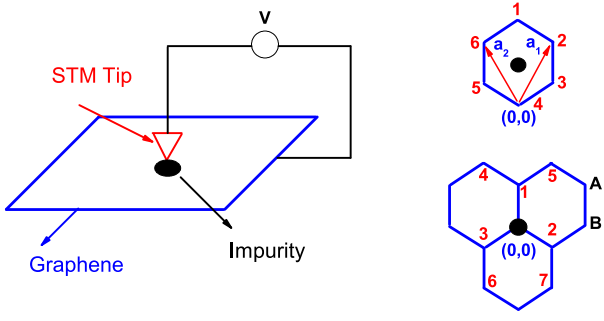}}
\caption{(Color online)Schematic experimental setup with the right
panel showing two possible positions (atop hexagon center and atop a
B site) of the impurity. The numbers denote nearest neighbor $A$ and
$B$ sublattice sites to the impurity. $a_{1(2)}= +(-) \sqrt{3}/2
\hat x + 3/2 \hat y$ [lattice spacing set to unity] are graphene
lattice vectors. The choice of coordinate center (0,0) are shown for
each case.} \label{fig1}
\end{figure}

\section{Computation of tunneling current}
\label{secfor}

The experimental situation for STM measurement is  schematically
represented in Fig.\ \ref{fig1}. The STM tip is placed atop the
impurity and the tunneling current ${\mathcal I}$ is measured as a
function of applied bias voltage $V$. The possible positions of the
impurity is shown in the right panel of Fig.\ \ref{fig1}. Such a
situation can be modeled by the well-known Anderson Hamiltonian
\cite{anderson1}. Here we incorporate the low-energy Dirac
quasiparticles of graphene in this Hamiltonian which is given by
\begin{eqnarray}
H &=& H_G + H_d+ H_{t} + H_{Gd} + H_{Gt} + H_{dt} \label{hamil1}
\end{eqnarray}
\begin{eqnarray}
H_G &=& \int_k \psi_s^{\beta \dagger}(\vec k)\Big [\hbar v_F (\tau_z
\sigma_x k_x + \sigma_y k_y) \nonumber\\
&& -E_F I \Big] \psi_s^{\beta}(\vec k) \label{hamilg} \\
 H_d &=& \sum_{s=\uparrow,\downarrow} \epsilon_d d_{s}^{\dagger} d_{s} + U
n_{\uparrow}n_{\downarrow}
\label{hamilimp} \\
H_{t} &=& \sum_{\nu} \Big[ \sum_{s=\uparrow \downarrow}
\epsilon_{t\nu} {\tilde t}_{\nu s}^{\dagger} {\tilde t}_{\nu s} +(\Delta_0 {\tilde t}_{\nu
\uparrow}^{\dagger} {\tilde t}_{-\nu \downarrow}^{\dagger}+ {\rm h.c} )\Big]
\label{hamiltip} \\
H_{Gd} &=& \sum_{\alpha=A,B} \int_k \left(V^0_{\alpha} (\vec k)
c_{\alpha,s}^{\beta}(\vec k) d_{s}^{\dagger} + {\rm h.c.} \right)
\label{graimp}\\
H_{dt}&=& \sum_{s=\uparrow,\downarrow;\nu} \left( W^0_{\nu} {\tilde t}_{\nu
s} d_{s}^{\dagger}+{\rm h.c.} \right). \label{imptip} \\
H_{Gt} &=& \sum_{\alpha=A,B;\nu} \int_k \left(U^0_{\alpha;\nu} (\vec
k) c_{\alpha,s}^{\beta}(\vec k) {\tilde t}_{\nu s}^{\dagger} + {\rm h.c.}
\right)\label{gratip}
\end{eqnarray}
Here $H_G$ is the Dirac Hamiltonian for the graphene electrons which
are described by the two component annihilation operator
\begin{eqnarray}
\psi^{\beta}_s (\vec k) &=& (c_{A s}^{\beta}(\vec k), c_{B
s}^{\beta}(\vec k))
\end{eqnarray}
belonging to the valley $\beta=K,K'$ and spin
$s=\uparrow,\downarrow$, $I$ is the identity matrix, $\tau$ and
$\sigma$ denote Pauli matrices in valley and pseudospin spaces,
$v_F$ is the Fermi velocity, and $\int_k \equiv \sum_{\beta= K,K'}
\sum_{s=\uparrow \downarrow} \int \frac{d^2k}{(2\pi)^2}$. $H_d$
denotes the impurity atom Hamiltonian with an on-site energy
$\epsilon_d$ and $U$ is the strength of on-site Hubbard interaction.
$H_t$ is the Hamiltonian for the superconducting ($\Delta_0\ne 0$)
or metallic ($\Delta_0=0$) tip electrons with on-site energy
$\epsilon_{t\nu}$, where $\nu$ signifies all quantum numbers (except
spin) associated with the tip electrons. The operators $d_{s}$ and
${\tilde t}_{\nu s}$ are the annihilation operators for the impurity and the
tip electrons. The Hamiltonians $H_{Gd}$, $H_{Gt}$, and $H_{dt}$
describe hopping between the graphene and the impurity electrons,
the graphene and the STM tip electrons, and the impurity and the STM
tip electrons, respectively. The corresponding parameters
$V^0_{\alpha}(\vec k)$, $U^0_{\alpha;\nu}(\vec k)$, and $W^0_{\nu}$
are taken to be independent of valley and spin indices of graphene
electrons but may depend on their sublattice index or pseudospin.
Note that the tunneling terms [Eq.\ \ref{graimp}] automatically take
into account potential scattering; such terms are generated once the
impurity degree of freedom is integrated out unless there is perfect
particle-hole symmetry ($E_F=0$).

The tunneling current for the present model is given by
\begin{eqnarray}
{\mathcal I}(t)&=&  e\langle dN_t/dt \rangle =ie \langle
[H,N_t]\rangle/\hbar,
\end{eqnarray}
where $N=\sum_{\nu s} {\tilde t}_{\nu s}^{\dagger} {\tilde t}_{\nu s}$ is the number
operator for the tip electrons. These commutators receive
contribution from $H_{dt}$ and $H_{Gt}$ in Eqs.\ (\ref{imptip}) and
(\ref{gratip})and can be written as
\begin{eqnarray}
{\mathcal I}(t) =  \frac{e}{\hbar} \Bigg[ \sum_{\sigma \nu}
\left(W_{\nu}^{0\,\ast} {\mathcal G}_{\sigma \nu}^{(2)<}(t) -
W_{\nu}^0
{\mathcal G}_{\nu \sigma}^{(2)<}(t) \right) \nonumber\\
+ \int_k \sum_{\sigma \nu} \left(U_{\nu}^{0\,\ast}(\vec k) G_{\sigma
\nu}^{(1)\, <}(t;\vec k) - U_{\nu}^0(\vec k) G_{\nu \sigma}^{(1)\,
<}(t;\vec k) \right)\Bigg] \label{cexp1}
\end{eqnarray}
where we define the standard Keldysh Green's function ${\mathcal
G}$ and $G$ as \cite{mahan1}
\begin{eqnarray}
G_{\sigma \nu}^{(1)\, <}(t;\vec k) &=& -i \langle {\tilde  t}_{\nu
\sigma}^{\dagger}(t)
\psi_{\sigma}(0;\vec k) \rangle \nonumber\\
G_{\nu \sigma}^{(1)\, <}(t;\vec k)&=&
-i \langle \psi_{\sigma}^{\dagger}(t;\vec k) {\tilde  t}_{\nu \sigma}(0) \rangle \nonumber\\
{\mathcal G}_{\sigma \nu}^{(2)<}(t) &=& -i \langle {\tilde  t}_{\nu
\sigma}^{\dagger}(t) d_{\sigma}(0) \rangle \nonumber\\
{\mathcal G}_{\nu \sigma}^{(2)<}(t)&=& -i \langle
d_{\sigma}^{\dagger}(t) {\tilde t}_{\nu \sigma}(0) \rangle
\label{gexp1}
\end{eqnarray}
These hybrid Green's functions (Eq.\ \ref{gexp1}) obey the usual
Keldysh relations. For example, ${\mathcal G}_{\sigma \nu}^{(2)\,<}$
and $ {\mathcal G}_{\sigma \nu}^{(2)\,>}$ can be expressed in terms
of the time ordered (${\mathcal G}_{\sigma \nu}^{(2)\,t}$),
anti-time ordered (${\mathcal G}_{\sigma,\nu}^{(2)\,\bar t}$),
retarded (${\mathcal G}_{\sigma,\nu}^{(2)\,R}$), and advanced
(${\mathcal G}_{\sigma,\nu}^{(2)\,A}$) Keldysh Green's functions as
\cite{mahan1}
\begin{eqnarray}
{\mathcal G}_{\sigma \nu}^{(2)\,t}+ {\mathcal G}_{\sigma
\nu}^{(2)\,\bar t} &=& {\mathcal G}_{\sigma \nu}^{(2)\,<} +
{\mathcal G}_{\sigma \nu}^{(2)\,>},
\nonumber\\
{\mathcal G}_{\sigma \nu}^{(2)\,R}- {\mathcal G}_{\sigma
\nu}^{(2)\,A} &=& {\mathcal G}_{\sigma \nu}^{(2)\,>} - {\mathcal
G}_{\sigma \nu}^{(2)\,<}. \label{kel1}
\end{eqnarray}
Similar relations hold for $G_{\sigma \nu}^{(1)\,<}(t;\vec k)$ and
$G_{\nu \sigma}^{(1)\,<}(t;\vec k)$.

Next, we expand the hybrid Green's functions $G_{\sigma \nu}^{(1)\,
<}(t;\vec k)$, $G_{\nu \sigma}^{(1)\, <}(t;\vec k)$, ${\mathcal
G}_{\sigma \nu}^{(2)<}(t)$ and ${\mathcal G}_{\nu \sigma}^{(2)<}(t)$
in perturbation series \cite{mahan1}. After some straightforward
algebra, one obtains, to first order in perturbation theory,
\begin{eqnarray}
G_{\sigma \nu}^{(1)\, <}(k) &=& \int_{k'} \sum_{\sigma' \nu'}
U_{\nu'}^0(\vec k') \Big[ g^t_{\nu'\sigma'; \nu \sigma}
{\mathcal G}_{\sigma,\sigma'}^{<}({\vec k},{\vec k}') \nonumber\\
&& -  g^<_{\nu' \sigma'; \nu \sigma}{\mathcal
G}_{\sigma,\sigma'}^{{\bar t} }({\vec k},{\vec k}') \Big]
+ \sum_{\sigma' \sigma} W_{\nu'}^{0}  \nonumber\\
&& \times \left[ g^t_{\nu'\sigma';\nu \sigma} G_{\sigma
\sigma'}^{h\, <}({\vec{k}}) - g^<_{\nu' \sigma'; \nu \sigma}
G_{\sigma \sigma'}^{h\,{\bar t}}({\vec k}) \right] \nonumber
\end{eqnarray}
\begin{eqnarray}
G_{\nu \sigma}^{(1)\, <}(\vec k) &=& \int_{k'} \sum_{\sigma'
\nu'} U_{\nu'}^{0\,\ast}(\vec k')
\Big[ g^<_{\nu \sigma;\nu' \sigma'}{\mathcal G}_{\sigma',\sigma}^{t}({\vec k},{\vec k}')\nonumber\\
&& - g^{\bar t}_{\nu\sigma;\nu' \sigma'}{\mathcal
G}_{\sigma',\sigma}^{<}({\vec k},{\vec k}')\Big]
+ \sum_{\sigma' \nu'} W_{\nu'}^{0\,\ast}  \nonumber\\
&& \times \left[ g^<_{\nu \sigma; \nu' \sigma'}
G_{\sigma'\sigma}^{h\,t}(\vec k) -
g^{\bar t}_{\nu \sigma;\nu' \sigma'} G_{\sigma';\sigma}^{h\,<}(\vec k) \right] \nonumber\\
{\mathcal G}_{\sigma \nu}^{(2)\, <} &=& \int_{k'} \sum_{\sigma'
\nu'} U_{\nu}^0 (\vec k')
\Big[ g^t_{\nu'\sigma'; \nu \sigma}G_{\sigma \sigma'}^{h\,<}({\vec k'}) \nonumber\\
&& - g^<_{\nu' \sigma';\nu \sigma} G_{\sigma \sigma'}^{(h\, {\bar t}
}({\vec k}') \Big]
+ \sum_{\sigma' \nu'} W_{\nu'}^0  \nonumber\\
&& \times \left[ g^t_{\nu'\sigma'; \nu \sigma} G_{\sigma
\sigma'}^{d\,<}-
g^<_{\nu' \sigma'; \nu \sigma} G_{\sigma \sigma'}^{d\,{\bar t}} \right] \nonumber\\
{\mathcal G}_{\nu \sigma}^{(2)\, <} &=& \int_{k'} \sum_{\sigma'
\nu'} U_{\nu'}^{0 \, \ast} (\vec k')
\Big[ g^<_{\nu \sigma;\nu' \sigma'} G_{\sigma',\sigma}^{h\,t}({\vec k}') \nonumber\\
&& - g^{\bar t}_{\nu \sigma; \nu' \sigma'}
G_{\sigma,\sigma'}^{h\,<}({\vec k}') \Big]
+ \sum_{\sigma' \nu'} W_{\nu'}^{0\,\ast}  \nonumber\\
&& \times \left[ g^<_{\nu \sigma; \nu' \sigma'}
G_{\sigma'\sigma}^{h\,t}({\vec k}') - g^{\bar t}_{\nu \sigma; \nu'
\sigma'} G_{\sigma' \sigma}^{h\,<}({\vec k}') \right] \label{green1}
\nonumber\\
\end{eqnarray}
where all the Green's functions appearing in Eq.\ (\ref{green1}) are at
the same time $t$ which we have not written out explicitly for
clarity.  In Eq.\ (\ref{green1}), $g_{\nu \sigma; \nu' \sigma'}^<(t)
=-i \langle {\tilde t}_{\nu \sigma}^{\dagger}(t) {\tilde t}_{\nu'
\sigma'}(0)\rangle$ denotes the Green's function for the tip electrons
which, in frequency space, is given by
\begin{eqnarray}
g_{\nu \sigma; \nu' \sigma'}^{<}(\omega) &=& 2\pi i f(\epsilon_{t
\nu}-\mu_t) \delta_{\nu \nu'} \delta_{\sigma \sigma'}
\delta(\omega-\epsilon_{t\nu}). \label{green2}
\end{eqnarray}
where $f(x)= 1/(1+ \exp\left[x/k_B T\right])$ denotes the
Fermi-Dirac distribution function at a temperature $T$, $\mu_t$ is
the chemical potential for the tip electrons, and $k_B$ is the
Boltzman constant. Similar expressions can be obtained for $g^t$ and
$g^{\bar t}$ using Eq.\ (\ref{kel1}) \cite{mahan1}. ${\mathcal
G}_{\sigma \sigma'}^{<}(t;{\vec k},{\vec k}')= -i \langle
\psi_{\sigma'}^{\dagger}(t;{\vec k}) \psi_{\sigma}(0;{\vec
k}')\rangle$ denotes the
Green's function of the Dirac electrons in
the presence of the impurity. The retarded and advanced components
of this Green function  which we shall need in subsequent analysis
can be written as
\begin{eqnarray}
{\mathcal G}_{\sigma \sigma'}^{R(A)}({\vec k},{\vec k}')=
\delta_{\sigma \sigma'} \delta({\vec k}- {\vec k}') {\mathcal
G}_{\sigma} ^{(0) R(A)}(\vec k) + \int_{k_1} \int_{k_2}
\sum_{\sigma_1,\sigma_2} \nonumber\\
\times V^{0\,\ast}(\vec k) V^{0}(\vec k') {\mathcal G}_{\sigma
\sigma_1}^{R(A)}({\vec k},{\vec k}_1) G_{\sigma_1 \sigma_2}^{d
R(A)}({\vec k}_1,{\vec k}_2) {\mathcal G}_{\sigma_2
\sigma'}^{R(A)}({\vec k}_2,{\vec k}') \label{green3} \nonumber\\
\end{eqnarray}
where again it is understood that all Green's functions are at a given
time $t$ and $G_{\sigma \sigma'}^{d R(A)}(t) = -i \langle
d_{\sigma}^{\dagger}(t) d_{\sigma'} (0)\rangle$ denotes the
retarded(advanced) Green's function of the interacting impurity
electrons. In frequency space, this Green's function is given by
\begin{eqnarray}
G_{\sigma \sigma'}^{d R(A)}(\omega)&=& \frac{\delta_{\sigma
\sigma'}}{\omega - \epsilon_d -{\rm
Re}[\Sigma_d(\omega)] -(+) i {\rm Im}[\Sigma_d(\omega)]} \nonumber\\
\label{impgreen}
\end{eqnarray}
where $\Sigma_d(\omega)$ denotes the self-energy of the impurity in
the absence of the tip. $\Sigma_d$ receives contributions from both
the on-site Hubbard interaction $U$ of the impurity electrons and
the coupling of the impurity to the Dirac electrons. Note that we
have neglected the effect of the STM tip while computing ${\mathcal
G}_{\sigma\sigma'} ^{R(A)}(t;{\vec k},{\vec k}')$ which is justified
as long as we restrict ourselves to linear-response theory. In Eq.\
(\ref{green3}), ${\mathcal G}_{\sigma} ^{(0) R(A)}(t;\vec k)$ denotes
the single-particle Green's function for the graphene electrons in the
absence of the impurity and the STM tip and is given, in frequency
space,  by
\begin{eqnarray}
{\mathcal G}_{\sigma}^{(0) R(A)} (\omega,{\vec k}) &=& \frac{(\omega
+E_F)I -\hbar v_F (\tau_z \sigma_x k_x + \sigma_y k_y)}{(\omega+
E_F)^2 -\hbar^2v_F^2 |{\vec k}|^2 -(+) i\eta} \nonumber\\
\label{grgreen}
\end{eqnarray}
Finally, the Green's function $G_{\sigma \sigma'}^{h\,<}(t;{\vec k})=
-i\langle d_{\sigma}^{\dagger}(t) \psi_{\sigma'} (0;\vec k)\rangle$
used in Eq.\ (\ref{green1}) is a hybrid
Green's function whose retarded
and advanced components are given, within first-order perturbation
theory, by
\begin{eqnarray}
G_{\sigma\sigma'}^{h\,R(A)}(t;{\vec k}) &=& \sum_{\sigma_1}
V^{0}(\vec k) {\mathcal G}_{\sigma\sigma_1} ^{(0) R(A)}(0;\vec k)
{\mathcal G}_{\sigma_1 \sigma'}^{d R(A)}(t)  \label{green4}
\end{eqnarray}

Next we follow Ref.\ \onlinecite{wingreen1} to substitute Eqs.\
(\ref{green1}) and (\ref{green2}) in Eq.\ (\ref{cexp1}) and approximate
the coupling functions to be independent of momentum: $U^{0}(\vec
k)\equiv U^0$, $W_{\nu}^0 \equiv W^0$, and $V^0(\vec k) \equiv V^0$.
Such an approximation is justified as long we restrict ourselves to
low applied voltages. With this approximation, after some algebra
involving Eqs.\ (\ref{cexp1})..(\ref{green4}), one obtains the
expression of the current
\begin{eqnarray}
{\mathcal I} = {\mathcal I}_0
\int_{-\infty}^{\infty}{d\omega}[f(\omega-eV)-f(\omega)]
\rho_t(\omega-eV) \Big[ \rho_G(\omega) \nonumber\\
\times |U^{0}|^2 + \frac{|B(\omega)|^2}{{\rm
Im}\Sigma_d(\omega)}\frac {|q(\omega)|^2-1+2{\rm Re}[q(\omega)]
\chi(\omega)}{(1+\chi^2(\omega))(1+\xi^2)}\Big ] \label{cur1}
\end{eqnarray}
where ${\mathcal I}_0 = 2e(1+\xi^2)/h$, $\rho_G(\epsilon)$ and
$\rho_t(\epsilon)$ are the graphene and STM tip electron DOS,
respectively, $\xi = |U^0_B|/|U^0_A|= |V^0_B|/|V^0_A|$ is the ratio
of coupling of the impurity to the electrons in $B$ and $A$ sites of
graphene with $U_A^0=U^0$ and $V_A^0=V^0$, and $\Sigma_d(\epsilon)$
is the impurity advanced self-energy in the absence of the tip. Here
$B(\epsilon)= V^0 U^0 I_2(\epsilon)$ and $q(\epsilon)$ and
$\chi(\epsilon)$ are given by
\begin{eqnarray}
q(\epsilon)&=& [W^0/U^0 + V^0 I_1(\epsilon)]/ [V^0 I_2(\epsilon)]
\nonumber\\
\chi(\epsilon)&=& \frac{\epsilon -\epsilon_d - {\rm Re}
\Sigma_d(\epsilon)}{{\rm Im} \Sigma_d(\epsilon)} \label{qeq}
\end{eqnarray}
where we have neglected the energy dependence of the coupling
functions assuming small applied voltages. In Eq.\ (\ref{qeq}),
$I_1(\epsilon)= (1+\xi^2) \sum_{k} {\rm Tr} \left[ {\rm
Re}\{{\mathcal G}^{(0)R}_{\sigma}(\epsilon,{\bf k})\}\right]$,
$I_2(\epsilon) = (1+\xi^2)\sum_k {\rm Tr} \left[{\rm Im}\{{\mathcal
G}^{(0) R}_{\sigma} (\epsilon,{\bf k})\}\right]$, and ${\rm Tr}$
denotes trace over Pauli matrices in pseudospin, valley and spin
spaces. Substituting Eq.\ (\ref{grgreen}) in Eq.\ (\ref{qeq}), we find
\cite{wingreen1, neto2}
\begin{eqnarray}
I_1(\epsilon) &=& -4(1+\xi^2)(\epsilon+E_F)\ln \left|1-
\Lambda^2/(\epsilon + E_F)^2 \right|/\Lambda^2 \nonumber\\
I_2(\epsilon) &=& 4(1+\xi^2) \pi |\epsilon+ E_F|
\theta(\Lambda-\epsilon - E_F ) /\Lambda^2. \label{i1i2}
\end{eqnarray}
where $\Lambda$ is the ultraviolet momentum cutoff and $\theta$ is
the Heaviside step function. Usually, in graphene, $\Lambda$ is
taken to be the energy at which the graphene bands start bending
rendering the low-energy Dirac theory inapplicable and can be
estimated to be $1-2$eV \cite{baskaran1}.

Equations\ (\ref{cur1}-\ref{i1i2}) constitute the central
results of this section and yields the tunneling current through the
STM tip within linear-response theory. We are going to analyze these
equations in the subsequent sections.

\section{Results}
\label{secresult}

\begin{figure}
\rotatebox{0}{\includegraphics*[width=0.9\linewidth]{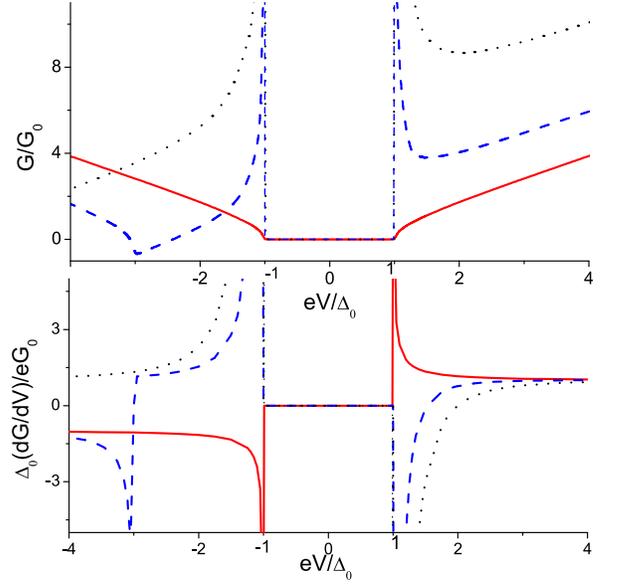}}
\caption{(Color online) Plot of the tunneling conductance $G$ and
its derivative $dG/dV$ as a function of the applied bias voltage
$eV/\Delta_0=-p$ for $r=0,2,6$ (red solid, blue dashed, and black
dotted lines), respectively. See text for details.} \label{fig2}
\end{figure}

In this section, we are going to analyze the tunneling conductance
($G(V)=d{\mathcal I}/dV$) as measured by the STM tip. First we
consider the case of a superconducting tip in the absence of any
impurity. In this case, the contribution to the conductance comes
from the first term of Eq.\ (\ref{cur1}). For {\it s}-wave superconducting
tips, one finds that the tunneling conductance ($G(V)=d{\mathcal
I}/dV$) for $E_F>0$ and at $T=0$ is given by (with $r=E_F/\Delta_0$,
$p=-eV/\Delta_0$)
\begin{eqnarray}
G &=& G_0 \Big[{\mathcal N}_t(p)|r| + \int_{p} {\rm Sgn} (z-p+r)
{\mathcal N}_t(z)dz \Big] \label{freegra1} \\
\frac{dG}{dV} &=& \frac{eG_0}{\Delta_0} \Big[{\mathcal N}_t(p)
-{\mathcal N}_t^{'}(p)|r| - 2 \theta(p-r) {\mathcal N}_t(p-r) \Big] \nonumber\\
\label{freegra2}
\end{eqnarray}
where $ G_0= 8 \pi^2 e^2|U^{0}|^2(1+\xi^2) \rho_{0t} \rho_0 \Delta_0
/h$, $\rho_G = \rho_0 |r-p|$, $\rho_t(r)=\rho_{0t} {\mathcal
N}_t(r)$, ${\mathcal N}_t (x)= |x|/\sqrt{x^2-1} \theta(|x|-1)$,
${\rm sgn}(x)$ denotes the signum function, $\rho_0= 6\sqrt{3}/(2 \pi
\hbar^2 v_F^2)$ (Ref. \
\onlinecite{neto1}) and $\rho_{0t}$ is the constant DOS of
the metallic tip. For graphene with $E_F=r=0$, $dG/dV \sim {\rm
Sgn}(V) {\mathcal N}_t(-V)$, {\it i.e.}, the tip DOS is given by the
derivative of the tunneling conductance. For large $E_F$ away from
the Dirac point, the first term of $G$ becomes large and reflects
the tip DOS. In between these extremes, when $E_F \sim eV$, neither
$G$ nor $dG/dV$ reflects the DOS. In this region, the signature of
the Dirac point appears through a cusp (discontinuity) in $G\,
(dG/dV)$ at $eV=-E_F-\Delta_0$ arising from the contribution of the
second (third) term in Eq.\ (\ref{freegra1}) [Eq.\ (\ref{freegra2})].
These features, shown in Fig.\ \ref{fig2}, distinguish such
graphene STM spectra with their conventional counterparts
\cite{stmsc}.

Next, we turn to the case of impurity-doped graphene and consider a
metallic tip with constant DOS. The contribution to the tunneling
conductance from the impurity (after subtracting the graphene
background) at $T=0$ [Eq.\ (\ref{i1i2})] is
\begin{eqnarray}
G_{\rm imp} &=& G_0^{'}\frac{|B(V)|^2}{{\rm Im}\Sigma_d(V)}\frac
{|q(V)|^2-1+2 {\rm Re} [q(V)]\chi(V)}{\Lambda[1+\chi^2(V)]},
\label{imptun}
\end{eqnarray}
where $G^{'}_0= 2e^2 \rho_{0t}\Lambda/h$. Such tunneling
conductances are known to have peak/antiresonace/dip feature at zero
bias for $|q| \gg 1/\simeq 1/\ll 1$ \cite{madhavan1}. In
conventional metals or earlier STM studies in graphene
\cite{peres1}, $U^0$ has been taken to be a fixed parameter
independent of the position of the impurity. However, as we show
here, the situation in graphene necessitates a closer attention. To
this end, we note that $U^0$ is proportional to the probability
amplitude of the Dirac quasiparticles in graphene to hop to the tip.
and its strength can be estimated using the well-known Bardeen
tunneling formula \cite {bardeen1}: $U^{0} \sim \int d^2 r
\left(\phi_{\nu}^{\dagger}(z)
\partial_z \Psi_G({\vec r},z) -\Psi_G^{
\dagger}({\vec r},z) \partial_z \phi_{\nu}(z)\right) \sim
\Psi_G({\vec r}_0,z_0)$, where the last similarity is obtained by a
careful evaluation of the surface integral $\int d^2 r$ over a
surface between the graphene and the tip parallel to the graphene
sheet\cite{tersoff1}, $({\vec r}_0,z_0)$ is the coordinate of the
tip center, $\phi_{\nu}(z)$ is tip electron wavefunction, and the
wave-function of the graphene electrons $ \Psi_G ({\vec r},z)$ around $K
(K')$ valley, can be written, within tight-binding approximation, as
\cite{mont1}
\begin{eqnarray}
\Psi_G ({\vec r},z) &=& \frac{1}{\sqrt{N}} \sum_{R_i^A} e^{ i
[\{\vec K (\vec K')+\vec {\delta k}\} \cdot {\vec R_i^A}]}
\Big[\varphi({\vec r}-{\vec R_i^A})
\nonumber\\
&& + e^{+(-)i \theta_k} \varphi({\vec r}-{\vec R_i^B}) \Big ] f(z).
\label{tightbinding}
\end{eqnarray}
Here $\theta_k = \arctan(k_y/k_x)$, $\vec {\delta k}$ is the Fermi
wave vector as measured from the Dirac points with $|\vec {\delta
k}| \ll |\vec K( \vec K')|$ for all $E_F$, $\varphi({\vec r})$ are
localized $p_z$ orbital wave functions, $N$ is a normalization
constant, $f(z)$ is a decaying function of $z$ with decay length set
by work function of graphene, and $ R_i^{A(B)} = n {\hat a}_1 + m
{\hat a}_2 ({\hat a}_2 - \hat y)$ with integers $n$ and $m$ denote
coordinates of the graphene lattice sites (Fig.\
\ref{fig1})\cite{mont1}. When the impurity and the STM tip is atop
the center of the hexagon, pseudospin symmetry necessitates
$\varphi({\vec r}_0-{\vec R_i^{A,B}})$ to be identical for all
neighboring $A$ and $B$ sublattice points $1-6$ surrounding the
impurity (Fig.\ \ref{fig1}). Consequently, the sum over lattice
vectors $R_i^A$ in Eq.\ (\ref{tightbinding}) reduces to a sum over the
phase factors $\exp (i [\{\vec K (\vec K')+ \vec {\delta k}\} \cdot
{\vec R_i^A}])$ for these lattice points. It is easy to check that
this sum vanishes for both Dirac points (when $|\vec {\delta
k}|=0$). Thus the only contribution to $\Psi_G({\vec r}_0,z_0)$
comes from the second and further neighbor sites for which the
amplitude of localized wave functions $\varphi({\vec r}_0-{\vec
R_i^{A/B}})$ are small. For finite $E_F$, ($\vec {\delta k} \ne 0$)
there is a finite but small contribution (${\rm O} (|\vec {\delta
k}|/|{\vec K|})$) to $\Psi_G({\vec r}_0,z_0)$ from the nearest-
neighbor sites. Thus $\Psi_G({\vec r}_0,z_0)$ and hence $U^0$ is
drastically reduced when the impurity is atop the hexagon center. In
this case, we expect $U^{0} \ll W^0$ and hence $|q| \gg 1$ [Eq.\
(\ref{qeq})] leading to a peaked spectra for all $E_F$. In contrast,
for the impurity atom atop a site, there is no such symmetry induced
cancellation and $\psi_G({\vec r}_0,z_0)$ receives maximal
contribution from the nearest graphene site directly below the tip.
Thus we expect $|U^0| \gg |W^0|$ (since it is easier for the tip
electrons to tunnel to delocalized graphene band than to a localized
impurity level) leading to $q \simeq I_1/I_2 \simeq
-\ln|1-\Lambda^2/(eV+E_F)^2|/\pi$. For large $|eV+E_F|$ and impurity
atop a site, $q \le 1$ leading to a dip or an antiresonance in
$G_{\rm imp}$ which is qualitatively distinct from the peaked
spectra for impurity atop the hexagon center. As $E_F \to 0$, $q$
diverges logarithmically for small $eV$. However, it can be shown
that in this regime $\chi$ shows a stronger linear divergence for
$eV \ne \epsilon_d$ which suppresses $G_{\rm imp}$. At
$eV=\epsilon_d$, the divergence of $\chi$ also becomes logarithmic
and we expect a peak of $G_{\rm imp}$. Note that these effects are
independent of $\Sigma_d$ and hence of the precise nature of the
impurity. Such an impurity position-dependent peak/dip structure of
$G_{\rm imp}$ has been observed for magnetic impurities in Ref.\
\onlinecite{hari1} for $E_F \gg eV$.
\begin{figure}
\rotatebox{0}{\includegraphics*[width=0.95\linewidth]{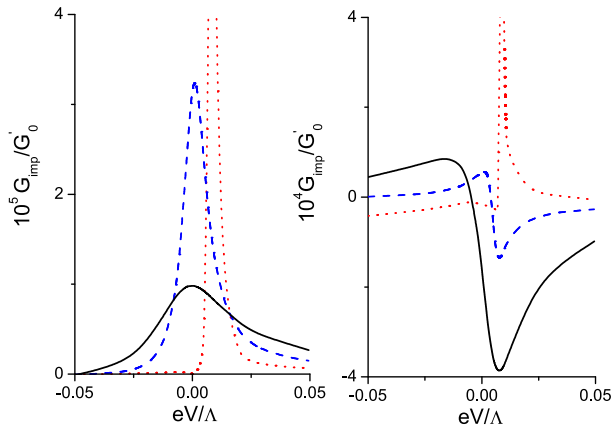}}
\caption{(Color online) Plot of $G_{\rm imp}$ as a function of $V$
for $|W^{0}/U^0|=0.05$ (right; impurity atop a site) and $2$ (left;
impurity atop hexagon center) for $E_F/\Lambda =0.3, 0.1, \,{\rm
and}\,0$ (black solid, blue dashed, and red dotted lines,
respectively). Plot parameters are $5U= V^{0}=0.05 \Lambda$, $W^0=
0.0005 \Lambda$, and $\epsilon_d =0$.} \label{fig3}
\end{figure}

To demonstrate this feature, we restrict ourselves to impurities
with small Hubbard $U$ and compute the self-energy of the impurity
electrons within a mean-field theory where $U n_{\sigma} n_{\bar
\sigma} = U \langle n_{\sigma} \rangle n_{\bar \sigma}$ leading to
spin-dependent on-site impurity energy $\epsilon_{\sigma} =
\epsilon_d +U \langle n_{\bar \sigma} \rangle$ \cite{neto2}. Using
Eqs.\ (\ref{hamil1}) and (\ref{graimp}), one then obtains the mean-field
advanced impurity Green's function ${\mathcal G}^{{\rm
imp}}_{\sigma}(\omega) = (\omega -
\epsilon_{\sigma}-\Sigma_d(\omega))^{-1}$ where the impurity
self-energy is given by $\Sigma_d(\omega) = |V^0|^2 (I_1+i I_2)$ and
mean-field self-consistency condition demands $n_{\sigma} = \int
d\omega/\pi {\rm Im} {\mathcal G}^{{\rm imp}}_{\sigma}(\omega)$.
Following Ref.\ \onlinecite{neto2}, we solve these equations to get
$\chi(\epsilon)$, and ${\rm Im} \Sigma_d(\epsilon)$ which can be
substituted in Eq.\ (\ref{imptun}) to obtain $G_{\rm imp}$. We note,
from Eqs.\ (\ref{qeq}) and (\ref{imptun}), that $G_{\rm imp}/G_0^{'}$
depends on the ratios $E_F/\Lambda$, $V^0/\Lambda$, and $W^0/U^0$
which can not be quantitatively determined from the Dirac-Anderson
model. We therefore treat them as parameters of the theory
\cite{neto1,neto2} and compute $G_{\rm imp}$ for their
representative values as shown in Fig.\ \ref{fig3}. In accordance
with earlier discussions, we find that for large $E_F/\Lambda=0.3$,
$G_{\rm imp}$ has qualitatively different features; for the impurity
at the center of the hexagon, it shows a peak (left panel) while for
that atop a site (right panel), it shows a dip. The change in
$G_{\rm imp}$ from a dip to a peak via an antiresonance as a
function of $E_F/\Lambda$ when the impurity is atop a site can be
seen from right panel of Fig.\ \ref{fig3}. In contrast, the left
panel always shows peak spectra.

\section{Conclusion}

\label{seccon}

In conclusion, we have shown that the tunneling conductance spectra
of both doped and undoped graphene have unconventional features not
discussed in earlier studies \cite{peres1}. In particular, the STM
spectra of doped graphene depend qualitatively on the position of
the impurity in the graphene matrix. This feature is demonstrated to
be a direct consequence of pseudopsin symmetry and Dirac nature of
graphene quasiparticles.

Further experimental verification of our work would involve
measuring tunneling conductance of doped and undoped graphene by
varying $E_F$. For undoped graphene with $E_F =0$, we propose to
measure the tunneling conductance spectra using a superconducting
tip and verify that $dG/dV \sim \rho_t {\rm sgn}(V)$. For small
$E_F>0$, where it is possible to access the regime $eV>E_F$ in
experiments, we predict a cusp (discontinuity) in $G\,(dG/dV)$ at
$eV=-E_F -\Delta_0$ as a signature of the Dirac point. The variation
in the shape of the spectra of impurity-doped graphene with impurity
atop a site may also be experimentally studied.

We also note that the theory of tunneling conductance derived here
should also be applicable to the impurity-doped Dirac electrons on
the surface of strong topological insulators with a single Dirac
cone \cite{topoin}. In this case, we expect to find that the STM
spectra should change from a dip to a peak through an anti-resonance
as the Fermi energy is tuned toward the Dirac point. This behavior
is qualitatively similar to that shown in the right panel of Fig.\
\ref{fig3}. However, such a controlled tuning of Fermi energy of
topological insulators seems to be experimentally more difficult
than graphene.

{\it Note added. } Recently, we came to
know of Ref.\ \onlinecite{uchoa1} with related results.

\section*{ ACKNOWLEDGMENTS}

KS thanks A. Castro-neto, V. N. Kotov, and H. Manoharan for
discussions.


\begin{thebibliography}{99}

\bibitem{neto1} A. H. Castro Neto, F. Guinea, N. M. R. Peres, K. S. Novoselov,
and A. K. Geim, Rev. Mod. Phys. {\bf 81}, 109 (2009); T. Ando, J.
Phys. Soc. Jpn. {\bf 74} 777 (2005).

\bibitem{shar1} V. P. Gusynin and S. G. Sharapov, Phys. Rev. Lett. {\bf 95},
146801 (2005); N. M. R Peres, F. Guinea, and A. H. Castro Neto, Phys.
Rev, B {\bf 73}, 125411 (2006); V. Lukose, R. Shankar and G.
Baskaran, Phys. Rev. Lett., {\bf 98} 116802 (2007).

\bibitem{nov2} K. S. Novoselov, A. K. Geim, S. V. Morozov, D. Jiang, M. I. Katsnelson,
I. V. Grigorieva, S. V. Dubonos, and A. A. Firsov,  Nature {\bf
438}, 197 (2005);  Y. Zhang, Y. -W. Tan, H. L. Stormer, and P. Kim,
Nature {\bf 438}, 201 (2005); K. S. Novoselov, E. McCann, S. V.
Morozov, V. I. Fal’ko, M. I. Katsnelson, U. Zeitler, D. Jiang, F.
Schedin, and A. K. Geim, Nat. Phys. {\bf 2} 177 (2006).

\bibitem{beenakker1} C. W. J. Beenakker, Phys. Rev. Lett.
{\bf 97}, 067007 (2006); M. Titov and C. W. J Beenakker, Phys. Rev. B
{\bf 74}, 041401(R) (2006).

\bibitem{sengupta1}  S. Bhattacharjee and K. Sengupta, Phys. Rev.
Lett. {\bf 97}, 217001 (2006); S. Bhattacharjee, M. Maiti and K.
Sengupta \prb {\bf 76}, 184514 (2007); M. Maiti and K. Sengupta,
\prb {\bf 76}, 054513 (2007).

\bibitem{baskaran1} K. Sengupta and G. Baskaran, Phys. Rev. B {\bf 77},
045417 (2008); M. Hentschel and F. Guinea, Phys. Rev. B {\bf 76},
115407 (2007)

\bibitem{neto2} B. Uchoa, V. N. Kotov, N. M. R. Peres and A. H.
Castro Neto, \prl {\bf 101}, 026805 (2008).

\bibitem{sensor1} F. Schedin, A. K. Geim, S. V. Morozov, E. W. Hill, P. Blake,
M. I. Katsnelson, and K. S. Novoselov, Nature Mater. {\bf 6},
652(2007).

\bibitem{hari1} H. Manoharan (private communication).

\bibitem{peres1} N. M. R. Peres, S-W. Tsai, J. E. Santos, and R. M. Ribeiro, Phys. Rev. B {\bf 79},
155442 (2009); H. Zhuang, Q. Shun, and X.C. Xie, EPL
{\bf 86}, 58004 (2009); P. S. Cornaglia, G. Usaj, and C. A.
Balseiro., \prl {\bf 102}, 046801 (2009);  N. M. R. Peres, L. Yang,
and S-W. Tsai, New J. Phys. {\bf 11}, 095007 (2009); O. Poplavskyy,
M. O. Goerbig, and C. Morais Smith, Phys. Rev. B {\bf 80}, 195414
(2009).

\bibitem{stmgra1} I. Brihuega, P. Mallet, C. Bena, S. Bose, C. Michaelis, L. Vitali, F. Varchon,
L. Magaud, K. Kern, and J. Y. Veuillen, \prl {\bf 101}, 206802
(2008).

\bibitem{davis1} T. Valla, A. V. Fedorov, Jinho Lee, J. C. Davis, and G. D. Gu, Science {\bf 314}, 1914
(2006); see, e.g., O. Fischer, M. Kugler, I. Maggio-Aprile, C. Berthod, and C. Renner,
Rev. Mod. Phys. {\bf 79}, 353 (2007).

\bibitem{madhavan1} U. Fano, Phys. Rev {\bf 124} 1866 (1961);
V. Madhavan, W. Chen, T. Jamneala, M. F. Crommie, and N. S. Wingreen, Science {\bf 280}, 567 (1998).

\bibitem{wingreen1} Y. Meir and N. S. Wingreen, \prl {\bf
68}, 2512 (1992); Y. Meir, N. S. Wingreen, and P.A. Lee, \prl {\bf
70}, 2601 (1993).

\bibitem{stmsc}
S. H. Pan, E. W. Hudson, and J. C. Davis, Appl. Phys. Lett. {\bf 73}, 2992 (1998); A. Kohen, Th. Proslier,
T. Cren, Y. Noat, W. Sacks, H. Berger, and D. Roditchev, Phys. Rev. Lett. {\bf 97}, 027001 (2006); I. Guillamon,
H. Suderow, S. Vieira, and P. Rodiere, Physica C {\bf 468}, 537 (2008).

\bibitem{anderson1} P.W. Anderson Phys. Rev. {\bf 124} 41 (1961).

\bibitem{mahan1} See for example, G.D. Mahan, {\it Many-Particle
Physics} (Plenum Press, New York, 1981).

\bibitem{bardeen1}J. Bardeen, \prl {\bf 6}, 57 (1961).

\bibitem{tersoff1} J. Tersoff and D. R. Hamann \prl {\bf 50}, 1998
(1983).

\bibitem{mont1} C. Bena and G. Montambaux, New J. Phys. {\bf 11}, 095003 (2009).

\bibitem{topoin} D. Hsieh, D. Qian, L. Wray, Y. Xia, Y. S. Hor, R. J. Cava, and M. Z. Hasan,
Nature {\bf 452}, 970 (2008); Y. Xia, D. Qian, D. Hsieh, L. Wray, A. Pal, H. Lin, A. Bansil,
D. Grauer, Y. S. Hor, R. J. Cava and  M. Z. Hasan, Nat. Phys. {\bf 5}, 398 (2009).

\bibitem{uchoa1} B. Uchoa, L. Yang, S-W. Tsai, N. M. R. Peres, and
A. H. Castro Neto, Phys. Rev. Lett.
{\bf 103}, 206804 (2009).

\end{thebibliography}
\end{document}